\documentclass[prl,twocolumn,superscriptaddress,floatfix,showpacs,amsmath,letterpaper]{revtex4}

\usepackage{graphicx} 
\usepackage{dcolumn} 
\usepackage{bm} 
\usepackage{epstopdf} 
\usepackage{color} 

\newcommand {\MM} [1] {\ensuremath{#1}}
\newcommand {\tsp} [1] {\ensuremath{\mskip #1\thinmuskip}}

\newcommand {\RM}  [1] {\ensuremath{\textrm{#1}}}

\newcommand {\Giga} {\RM{G}}

\newcommand {\Volt} {\RM{V}}

\newcommand {\eV} {\tsp{0.1}\RM{e}\tsp{-0.3}\Volt}
\newcommand {\GeV} {\Giga\eV}

\newcommand {\unit} [1] {\MM{\tsp{1.5} #1}}

\newcommand {\capsword} [1] {\textsc{#1}}

\newcommand {\STAR} {\capsword{STAR}}

\newcommand {\IR} {\capsword{IR}}

\newcommand {\BEMC} {\capsword{BEMC}}
\newcommand {\EEMC} {\capsword{EEMC}}

\newcommand {\TPC} {\capsword{TPC}}

\begin{document}

\title{Measurement of the parity-violating longitudinal single-spin asymmetry for $W^{\pm}$ boson production in polarized proton-proton collisions at $\sqrt{s} = 500\,$GeV}


\affiliation{Argonne National Laboratory, Argonne, Illinois 60439, USA}
\affiliation{Brookhaven National Laboratory, Upton, New York 11973, USA}
\affiliation{University of California, Berkeley, California 94720, USA}
\affiliation{University of California, Davis, California 95616, USA}
\affiliation{University of California, Los Angeles, California 90095, USA}
\affiliation{Universidade Estadual de Campinas, Sao Paulo, Brazil}
\affiliation{University of Illinois at Chicago, Chicago, Illinois 60607, USA}
\affiliation{Creighton University, Omaha, Nebraska 68178, USA}
\affiliation{Czech Technical University in Prague, FNSPE, Prague, 115 19, Czech Republic}
\affiliation{Nuclear Physics Institute AS CR, 250 68 \v{R}e\v{z}/Prague, Czech Republic}
\affiliation{University of Frankfurt, Frankfurt, Germany}
\affiliation{Institute of Physics, Bhubaneswar 751005, India}
\affiliation{Indian Institute of Technology, Mumbai, India}
\affiliation{Indiana University, Bloomington, Indiana 47408, USA}
\affiliation{Alikhanov Institute for Theoretical and Experimental Physics, Moscow, Russia}
\affiliation{University of Jammu, Jammu 180001, India}
\affiliation{Joint Institute for Nuclear Research, Dubna, 141 980, Russia}
\affiliation{Kent State University, Kent, Ohio 44242, USA}
\affiliation{University of Kentucky, Lexington, Kentucky, 40506-0055, USA}
\affiliation{Institute of Modern Physics, Lanzhou, China}
\affiliation{Lawrence Berkeley National Laboratory, Berkeley, California 94720, USA}
\affiliation{Massachusetts Institute of Technology, Cambridge, MA 02139-4307, USA}
\affiliation{Max-Planck-Institut f\"ur Physik, Munich, Germany}
\affiliation{Michigan State University, East Lansing, Michigan 48824, USA}
\affiliation{Moscow Engineering Physics Institute, Moscow Russia}
\affiliation{NIKHEF and Utrecht University, Amsterdam, The Netherlands}
\affiliation{Ohio State University, Columbus, Ohio 43210, USA}
\affiliation{Old Dominion University, Norfolk, VA, 23529, USA}
\affiliation{Panjab University, Chandigarh 160014, India}
\affiliation{Pennsylvania State University, University Park, Pennsylvania 16802, USA}
\affiliation{Institute of High Energy Physics, Protvino, Russia}
\affiliation{Purdue University, West Lafayette, Indiana 47907, USA}
\affiliation{Pusan National University, Pusan, Republic of Korea}
\affiliation{University of Rajasthan, Jaipur 302004, India}
\affiliation{Rice University, Houston, Texas 77251, USA}
\affiliation{Universidade de Sao Paulo, Sao Paulo, Brazil}
\affiliation{University of Science \& Technology of China, Hefei 230026, China}
\affiliation{Shandong University, Jinan, Shandong 250100, China}
\affiliation{Shanghai Institute of Applied Physics, Shanghai 201800, China}
\affiliation{SUBATECH, Nantes, France}
\affiliation{Texas A\&M University, College Station, Texas 77843, USA}
\affiliation{University of Texas, Austin, Texas 78712, USA}
\affiliation{Tsinghua University, Beijing 100084, China}
\affiliation{United States Naval Academy, Annapolis, MD 21402, USA}
\affiliation{Valparaiso University, Valparaiso, Indiana 46383, USA}
\affiliation{Variable Energy Cyclotron Centre, Kolkata 700064, India}
\affiliation{Warsaw University of Technology, Warsaw, Poland}
\affiliation{University of Washington, Seattle, Washington 98195, USA}
\affiliation{Wayne State University, Detroit, Michigan 48201, USA}
\affiliation{Institute of Particle Physics, CCNU (HZNU), Wuhan 430079, China}
\affiliation{Yale University, New Haven, Connecticut 06520, USA}
\affiliation{University of Zagreb, Zagreb, HR-10002, Croatia}

\author{M.~M.~Aggarwal}\affiliation{Panjab University, Chandigarh 160014, India}
\author{Z.~Ahammed}\affiliation{Lawrence Berkeley National Laboratory, Berkeley, California 94720, USA}
\author{A.~V.~Alakhverdyants}\affiliation{Joint Institute for Nuclear Research, Dubna, 141 980, Russia}
\author{I.~Alekseev}\affiliation{Alikhanov Institute for Theoretical and Experimental Physics, Moscow, Russia}
\author{J.~Alford}\affiliation{Kent State University, Kent, Ohio 44242, USA}
\author{B.~D.~Anderson}\affiliation{Kent State University, Kent, Ohio 44242, USA}
\author{C.~D.~Anson}\affiliation{Ohio State University, Columbus, Ohio 43210, USA}
\author{D.~Arkhipkin}\affiliation{Brookhaven National Laboratory, Upton, New York 11973, USA}
\author{G.~S.~Averichev}\affiliation{Joint Institute for Nuclear Research, Dubna, 141 980, Russia}
\author{J.~Balewski}\affiliation{Massachusetts Institute of Technology, Cambridge, MA 02139-4307, USA}
% ...

\collaboration{STAR Collaboration}\noaffiliation

\begin{abstract}
We report the first measurement of the parity violating single-spin asymmetries for midrapidity decay positrons and electrons
from $W^{+}$ and $W^{-}$ boson production in longitudinally polarized proton-proton collisions at $\sqrt{s}=500\,$GeV by the STAR 
experiment at RHIC. The measured asymmetries, $A^{W^+}_{L}$$=-0.27\pm 0.10\;({\rm stat.})\pm 0.02\;({\rm syst.}) \pm 0.03\;({\rm norm.})$
and $A^{W^-}_{L}$$=0.14\pm 0.19\;({\rm stat.})\pm 0.02 \;({\rm syst.})\pm 0.01\;({\rm norm.})$, are consistent with theory predictions, 
which are large and of opposite sign. These predictions are based on polarized quark and antiquark distribution functions constrained by 
polarized DIS measurements.
\end{abstract}



\maketitle


Understanding the spin structure of the nucleon remains a fundamental challenge in Quantum Chromodynamics (QCD). 
Experimentally, polarized deep-inelastic scattering 
(pDIS) measurements have shown that the quark spins account for only $\approx 33\%$ of the proton 
spin \cite{Bass:2009dr}. Semi-inclusive pDIS measurements \cite{Adeva:1997qz, Airapetian:2004zf, Alekseev:2007vi} 
are sensitive to the quark and antiquark spin contributions separated by 
flavor \cite{deFlorian:2008mr, deFlorian:2009vb}. They rely on a quantitative understanding of the fragmentation of 
quarks and antiquarks into observable final-state hadrons. 
While the sum of the contributions from quark and antiquark parton distribution functions (PDFs) 
of the same flavor is well constrained, the uncertainties in the polarized antiquark 
PDFs separated by flavor remain relatively large 
\cite{deFlorian:2008mr, deFlorian:2009vb}.


High-energy polarized proton collisions at $\sqrt{s}=200\,$GeV and $\sqrt{s}=500\,$GeV at RHIC provide a unique way to probe
the proton spin structure and dynamics using hard scattering processes \cite{Bunce:2000uv}. 
The production of $W^{\pm}$ bosons at $\sqrt{s}=500\,$GeV
provides an ideal tool to study the spin-flavor structure of sea quarks inside the proton. 
$W^{+(-)}$ bosons are dominantly produced through $u+\bar{d}$ $(d+\bar{u})$ interactions and can be detected through
their leptonic decays \cite{Bourrely:1993dd}. 
Quark and antiquark polarized PDFs are probed directly in calculable 
leptonic $W$ decays at large
scales set by the mass of the $W$ boson. 
The production of $W$ bosons in polarized proton collisions allows for the observation of purely 
weak interactions, giving rise to large parity-violating longitudinal single-spin asymmetries.
A theoretical framework has been developed to describe inclusive lepton 
production, $\vec{p}+p \rightarrow W^{\pm}+X \rightarrow l^{\pm}+X$, that can be directly compared with experimental
measurements using constraints on the transverse energy and pseudorapidity of the final-state 
leptons \cite{Nadolsky:2003ga, deFlorian:2010aa}. 

\begin{figure}[t]
\includegraphics{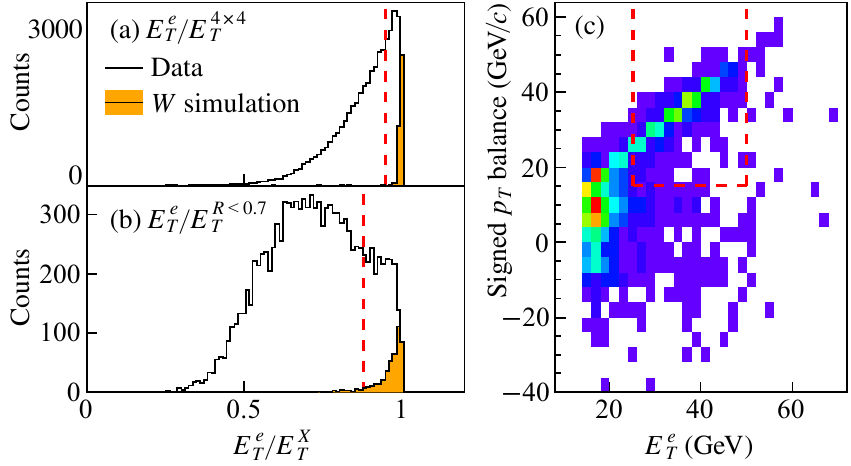}
\caption{{\it (a) Ratios of $E_{T}^{e}$ with respect to the $4\times 4$ BEMC $E_{T}$ sum, $E_{T}^{e}/E_{T}^{X=4 \times 4}$, (b) 
the near-cone BEMC, EEMC and TPC $E_{T}$ sum, $E_{T}^{e}/E_{T}^{X=R<0.7}$, 
and (c) correlation of the signed $p_{T}$ balance variable 
and $E_{T}^{e}$. MC shape 
distributions (arbitrary normalization) are shown in (a) and (b) for $W^\pm\rightarrow e^\pm+X$ as filled histograms in 
comparison to both data distributions.}}
\label{fig_Cuts}
\end{figure}
 

In this letter, we report the first measurement of the parity violating single-spin asymmetries for midrapidity decay positrons and electrons 
from $W^{+}$ and $W^{-}$ boson production in longitudinally polarized $\vec{p}+p$ collisions at $\sqrt{s}=500\,$GeV by the 
STAR experiment at RHIC. The asymmetry is defined as $A_{L} \equiv (\sigma^{+} - \sigma^{-})/(\sigma^{+} + \sigma^{-})$,
where $\sigma^{+(-)}$ is the cross section when the helicity of the polarized proton beam is positive (negative).


The STAR detector systems \cite{Ackermann:2002ad} used in this measurement are the Time Projection Chamber \cite{Anderson:2003ur} (\TPC) 
and the Barrel \cite{Beddo:2002zx} and Endcap \cite{Allgower:2002zy} Electromagnetic Calorimeters (\BEMC, \EEMC).
The TPC provides tracking for charged particles in a $0.5\,$T solenoidal magnetic field for 
pseudorapidities $|\eta|<1.3$ with full azimuthal coverage. The BEMC and EEMC 
are lead-scintillator sampling calorimeters providing full azimuthal coverage for
$|\eta|<1$ and $1.09<\eta<2$, respectively.


The data analyzed in this letter were collected in 2009 with colliding polarized proton beams at 
$\sqrt{s}=500\unit{\GeV}$ and an average luminosity of $55 \times 10^{30}\unit{\mathrm{cm}^{-2}\,\mathrm{s}^{-1}}$\tsp{-0.5}.
%
%
The polarization of each beam was measured 
using Coulomb-Nuclear Interference proton-carbon polarimeters~\cite {Nakagawa:2007zza}, 
which were calibrated using a polarized hydrogen gas-jet target~\cite {Makdisi:2007zz}\@. 
Longitudinal polarization of proton beams in the \STAR\ interaction region (\IR) 
was achieved by spin rotator magnets upstream and downstream of the \IR\ that changed the proton spin orientation from its stable vertical direction to 
longitudinal.  
Non-longitudinal beam polarization components were continuously monitored with a local polarimeter system at \STAR\ based
on the Zero-Degree Calorimeters with an upper limit on the relative contribution of $15\%$ for both polarized proton beams.   
The longitudinal beam polarizations averaged over all runs 
were $P_1 = 0.38$ and $P_2 = 0.40$ with correlated relative uncertainties 
of 8.3\% and 12.1\%, respectively. 
Their sum $P_1+P_2=0.78$ is used in the analysis and has a relative uncertainty 
of 9.2\%.


Positrons ($e^{+}$) and electrons ($e^{-}$) from $W^{+}$ and $W^{-}$ boson production with $|\eta_{e}|<1$ are 
selected for this analysis.
High-$p_{T}$ $e^{\pm}$ are charge-separated using the STAR TPC.
The BEMC is used to measure the transverse energy $E^{e}_{T}$ of $e^{+}$ and $e^{-}$. 
The suppression of the QCD background is achieved with the TPC, BEMC, and EEMC.  

The selection of $W$ candidate events is based on kinematic and topological differences between 
leptonic $W^{\pm}$ decays and QCD background events. Events from $W^\pm$ decays contain a nearly isolated 
$e^{\pm}$ with a neutrino in the opposite direction in azimuth. The neutrino escapes detection leading 
to a large missing energy.
Such events exhibit a large imbalance in the
vector $p_{T}$ sum of all reconstructed final-state objects. In contrast, QCD events, e.g. dijet events, are characterized 
by a small magnitude of this vector sum imbalance.

Candidate $W$ events were selected online by  
a two-step energy requirement in the BEMC. Electrons or positrons from $W$ production at midrapidity are characterized by
large $E_{T}$ peaked at $\approx M_{W}/2$ (Jacobian peak). At the hardware trigger level, a high tower calorimetric trigger
condition required $E_{T}>7.3\,$GeV in a single BEMC tower. At the software trigger level, 
a dedicated trigger algorithm searched for a seed tower of $E_{T}>5\,$GeV and computed all
four possible combinations of the $2 \times 2$ tower cluster $E_{T}$ sums and required at least one
to be above $13\,$GeV. A total 
of $1.4\times 10^{6}$ events were recorded for a data sample of $12\,$pb$^{-1}$. 
A Vernier scan was used to determine the absolute
luminosity \cite{SvdeMeer1968}.



\begin{figure}[t]
\includegraphics{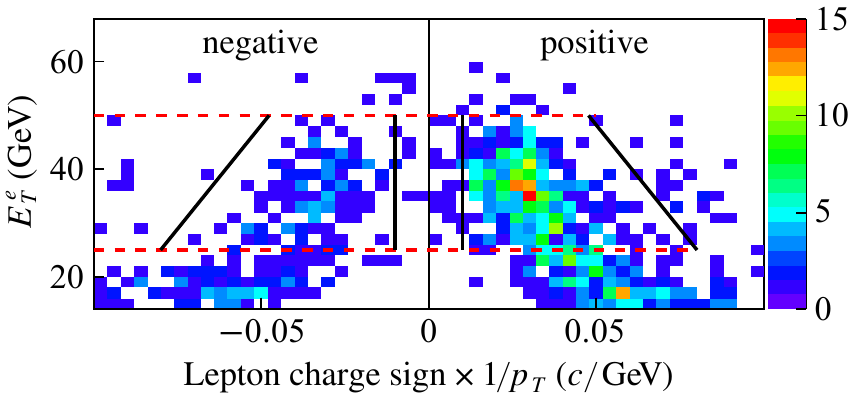}
\caption{{\it $E^{e}_{T}$ as a function of the ratio of the TPC reconstructed charge sign to the transverse momentum $p_{T}$. The black-solid and red-dashed lines 
indicate the selected kinematic region used for the asymmetry analysis.}} 
\label{fig_QvPT}
\end{figure}

An electron or positron candidate is defined to be any TPC track with $p_T>10\,$GeV$/c$ that is associated with a primary vertex 
with $|z|<100\,$cm, where $z$ is measured along the beam direction.
A $2\times 2$ BEMC tower cluster $E_{T}$ sum $E_{T}^{e}$, whose centroid is within 
$7\,$cm of the projected TPC track,
is required to be larger than $15\,$GeV. 
The excess BEMC $E_{T}$ sum in a $4 \times 4$ tower cluster centered around the 
$2\times 2$ tower cluster is required to be below $5\%$, as indicated by the vertical dashed line in 
Fig.~\ref{fig_Cuts} $({\rm a})$. 
A cone, referred
to as the near-side cone, is formed around the $e^{\pm}$ candidate with a radius $R=0.7$ in $\eta$-$\phi$ space. 
The excess BEMC, EEMC, and TPC $E_T$ sum in this cone is required to be less than $12\%$ of the $2\times 2$ cluster $E_T$, 
as shown in Fig.~\ref{fig_Cuts} $({\rm b})$ by the vertical dashed line. \textsc{pythia} 6.205 \cite{Sjostrand:2000wi} Monte-Carlo (MC) shape distributions 
(arbitrary normalization) for $W^\pm\rightarrow e^\pm+\nu$ passed through the \textsc{geant} \cite{Brun:1978fy} model of the 
STAR detector are shown in Figs.~\ref{fig_Cuts} (a) and~\ref{fig_Cuts} (b) as filled histograms motivating both ratio cuts.
The missing energy requirement is enforced by a cut on the $p_{T}$ balance vector, defined as the vector sum of the $e^{\pm}$ 
candidate
$p_{T}$ and the $p_{T}$ vectors of all reconstructed jets, 
where the jet thrust axis is required to be outside the near-side cone. Jets are reconstructed
using a standard mid-point cone algorithm used in STAR jet measurements \cite{Abelev:2006uq}
based on the TPC, BEMC, and EEMC.
A scalar signed $p_{T}$ balance variable is formed, given by the magnitude of the $p_{T}$ balance vector and 
the sign of the dot-product of the $p_{T}$ balance vector and the electron $p_{T}$ vector. This quantity is required to be 
larger than $15\,$GeV$/c$. The correlation of the signed $p_{T}$ balance variable and $E_{T}^{e}$ is shown in Fig.~\ref{fig_Cuts} $({\rm c})$. 
The range for accepted $W$ candidate events is marked by red dashed lines. The lower cut is chosen to suppress the contribution of 
background events
whereas the upper cut is mainly applied to ensure proper charge sign reconstruction.
Background events from $Z^{0}\rightarrow e^{+}e^{-}$ decays are suppressed by rejecting events with an additional 
electron/positron-like $2 \times 2$  
cluster in the reconstructed jet where the $E^{2\times 2}_{T}>p_{T}^{\rm jet}/2$ and the invariant mass of the two electron/positron-like 
clusters is within $70$ to $140\,$GeV$/c^{2}$. This avoids $Z^{0}$ contamination in the data-driven QCD background described below.


Figure~\ref{fig_QvPT} shows $E^{e}_{T}$ as a function of the ratio of the TPC reconstructed charge sign to the transverse momentum $p_{T}$
for electron and positron candidates that pass all the cuts described above. 
Two well-separated regions for positive (negative) charges are visible, identifying the $W^{+(-)}$ 
candidate events up to $E_{T}^{e} \approx 50\,$GeV.
The range of $E^{e}_{T}$ for accepted $W$ candidate events, $25<E^{e}_{T}<50\,$GeV, is marked by red dashed lines.
Entries outside the black solid lines in Fig.~\ref{fig_QvPT} were rejected due to false track reconstruction. 

\begin{figure}[t]
\includegraphics{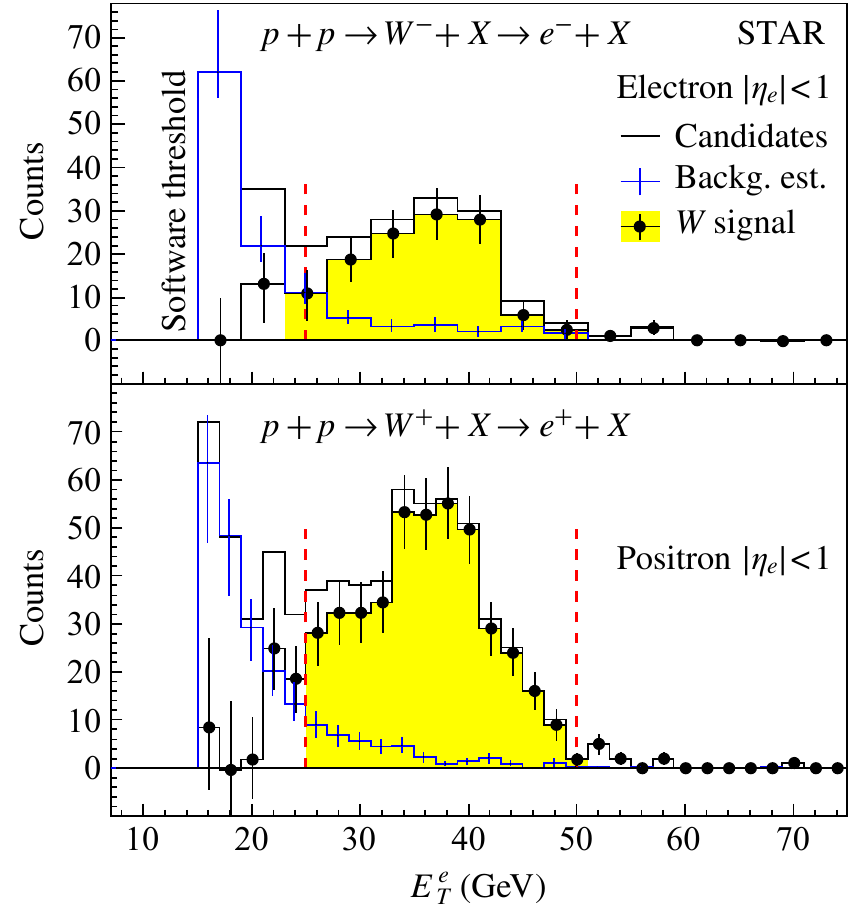}
\caption{{\it $E^{e}_{T}$ for $W^{+}$ (bottom) and $W^{-}$ (top) events showing the candidate histograms in black, the full background estimates in blue and
the signal distributions in yellow.}}
\label{fig_j_peak}
\end{figure}

Figure~\ref{fig_j_peak} presents the charge-separated lepton $E^{e}_{T}$ distributions based on the 
selection criteria given above. $W$ candidate events are shown as the black histograms, where the characteristic Jacobian peak can be seen at $\approx M_{W}/2$. 
The total number of candidate events for $W^{+(-)}$ is $462\,(139)$ for $25<E^{e}_{T}<50\,$GeV indicated
by vertical red dashed lines in Fig.~\ref{fig_j_peak}. 
The number of background events was estimated through a combination of \textsc{pythia} 6.205 \cite{Sjostrand:2000wi} MC simulations
and a data-driven procedure.  
The $e^{+(-)}$ background from $W^{+(-)}$ boson induced $\tau^{+(-)}$ decays and $Z^{0}\rightarrow e^{+}+e^{-}$ decays was estimated
using MC simulations to be $10.4\pm 2.8$ ($0.7\pm 0.7$) events and $8.5\pm 2.0$ events (identical for both $e^{+(-)}$), respectively.
The remaining background is mostly due to QCD dijet events where one of the jets missed the 
STAR acceptance. We have developed a data-driven procedure to evaluate this type of background. 
We excluded the EEMC ($1.09<\eta<2$) as an active detector in our analysis to estimate the background due to missing 
calorimeter coverage for $-2<\eta<-1.09$. 
The background contribution due to missing calorimeter coverage along with 
$\tau$ and $Z^{0}$ background contributions have been 
subtracted from both $W^{+(-)}$ $E_{T}^{e}$ distributions. 
The remaining background, 
presumably due to missing jets outside the STAR $|\eta|<2$ window, is evaluated 
based on an extrapolation from the region of $E^{e}_{T}<19\,$GeV in both $W^{+(-)}$ $E_{T}^{e}$ 
distributions. The shape is determined from the $E_{T}^{e}$ distribution in events previously rejected
as background with systematic variations of the signed $p_{T}$ balance cut below $15\,$GeV$/c$. 
This shape $E_{T}^{e}$ distribution is normalized to both $W^{+(-)}$ $E_{T}^{e}$ distributions for
$E^{e}_{T}<19\,$GeV.
The total number of background events for $e^{+(-)}$ is $39\pm 9$ ($23\pm 6$) for $25<E^{e}_{T}<50\,$GeV
shown in Fig.~\ref{fig_j_peak} 
as the blue line. The errors on the total background are mostly from the data-driven background events.

The leptonic asymmetry from $W^{\pm}$ decay, $A^{W^{\pm}}_{L}$, was obtained from:
\begin {equation}
\label {all_def}
A^{W^{\pm}}_{L} = \frac{1}{\beta^{\pm}}\frac{2}{P_1 +P_2}\frac{R_{++}N^{W^\pm}_{++} - R_{--}N^{W^\pm}_{--} \raisebox{-1ex}{} }{\Sigma_{i} R_{i}N^{W^\pm}_{i} \raisebox{1ex}{} }-\frac{\alpha^{\pm}}{\beta^{\pm}}
\end {equation}
where $P_{1,2}$ are the mean polarizations, $N^{W^\pm}_{i}$ are $W^{\pm}$ candidate yields for all four beam 
helicity configurations $i=\{++,+-,-+,--\}$, and $R_{i}$ are the
respective relative luminosities. 
The longitudinal single-spin asymmetry $A_{L}$ for $Z^{0}$ bosons has been 
estimated using a full next-to-leading (NLO) order framework \cite{deFlorian:2010aa}. With the $W^{\pm}$ selection criteria we 
estimated the $Z^{0}$ asymmetry to be $A_{L}^{Z}=-0.06$. This value has been used to determine the polarized
background contribution $\alpha^{+(-)}=-0.002\pm 0.001\, (-0.005\pm 0.002)$. 
The unpolarized background correction for $W^{\pm}$ candidate events 
is $\beta^{+(-)}=0.938\pm 0.017\,(0.838\pm 0.032)$.
This dilution factor is due to background events passing all $W$ selection cuts and is determined by $\beta=S/(S+B)$, where $S$ ($B$) is 
the number of signal (background) events for $25<E^{e}_{T}<50\,$GeV.


The relative luminosities $R_{i}=\sum_{k}M_{k}/(4M_{i})$ are determined from the ratios of yields $M_i$ of QCD events, 
for which parity conservation is expected. The $M_i$ are statistically independent 
from $N^{W^\pm}_{i}$ because the isolation cut on the $2\times 2$ / $4\times 4$ tower $E_{T}$ sum, shown in Fig.~\ref{fig_Cuts}, 
was reversed for those events. 
Additionally, an upper limit of $20\,$GeV was set on $E_{T}^{e}$.



\begin{figure}[t]
\includegraphics{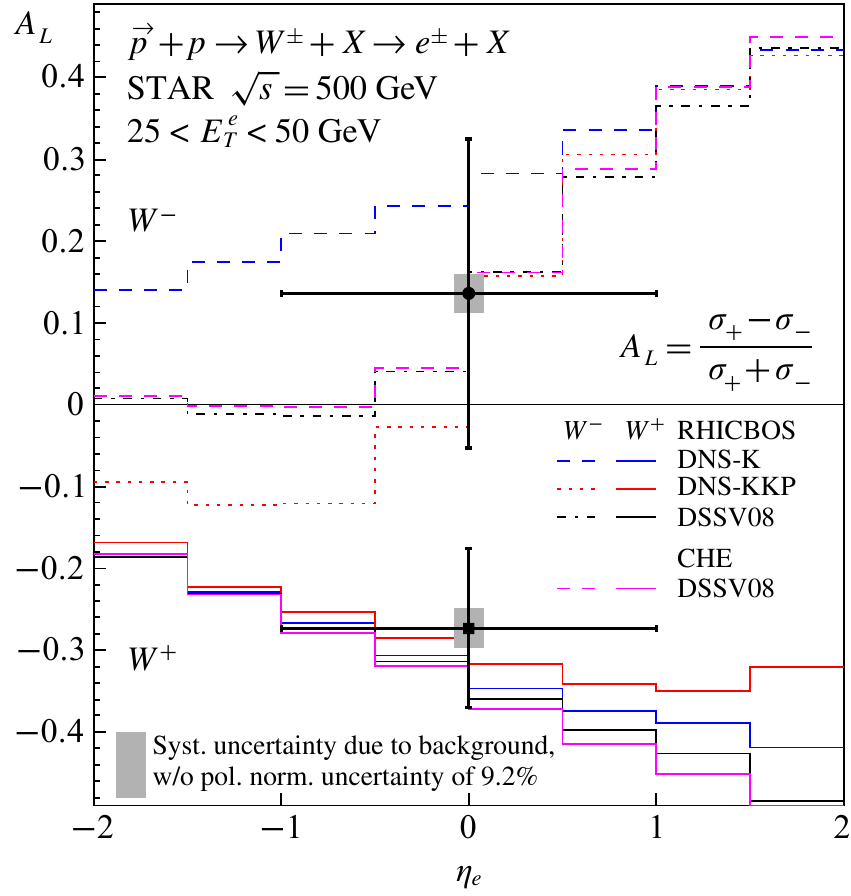}
\caption{{\it Longitudinal single-spin asymmetry, $A_{L}$, for $W^{\pm}$ 
events as a function of the leptonic pseudorapidity, $\eta_{e}$, for $25<E^{e}_{T}<50\,$GeV in comparison to theory predictions (See text for details).}}
\label{fig_wal}
\end{figure}

Figure~\ref{fig_wal} shows the measured leptonic asymmetries $A^{W^+}_{L}=-0.27\pm 0.10\;({\rm stat.})\pm 0.02\;({\rm syst.})$
and $A^{W^-}_{L}=0.14\pm 0.19\;({\rm stat.})\pm 0.02 \;({\rm syst.})$
for $|\eta_{e}|<1$ and $25<E^{e}_{T}<50\,$GeV. The vertical black error bars include only the statistical uncertainties. The systematic
uncertainties are indicated as grey bands. The statistical uncertainties dominate over the systematic uncertainties. 
%
%
The asymmetry $A_{L}$ observed in statistically independent samples of QCD dominated events was found to be 
$0.04 \pm 0.03$ ($0.00 \pm 0.04$) for positive (negative) charged tracks and is consistent with zero.
We assumed the experimental limit on the polarized background $A_{L}$ to be $0.02$ as a systematic uncertainty of $A^{W^{\pm}}_{L}$.
This limit on polarized background and the uncertainty in unpolarized background dilution 
have been added in quadrature to account for the total systematic uncertainty of $A^{W^{\pm}}_{L}$.
The normalization uncertainty of the measured asymmetries 
due to the uncertainty for the polarization sum $P_{1}+P_{2}$ is $0.03$ ($0.01$) for $A^{W^{+(-)}}_{L}$. 
The normalization uncertainty is of similar size as the systematic uncertainty of the asymmetry measurement.


In Fig.~\ref{fig_wal}, the measured asymmetries are compared to predictions based on full resummed (\textsc{rhicbos}) 
\cite{Nadolsky:2003ga} and NLO (\textsc{che}) \cite{deFlorian:2010aa} calculations. The \textsc{che} calculations use the DSSV08 
polarized PDFs [5], whereas the \textsc{rhicbos} 
calculations are shown in addition for the older DNS-K and DNS-KKP \cite{deFlorian:2005mw} PDFs.  The \textsc{che} 
and \textsc{rhicbos} 
results are in 
good agreement. The range spanned by the DNS-K and DNS-KKP distributions for $\Delta \bar{d}$ and $\Delta \bar{u}$ coincides, 
approximately, with the corresponding DSSV08 uncertainty estimates \cite{deFlorian:2008mr, deFlorian:2009vb}.
%
%
The spread of predictions for $A^{W^{+(-)}}_{L}$ is largest at forward (backward) $\eta_{e}$ and is strongly correlated to the 
one found for the $\bar{d}$ ($\bar{u}$) polarized PDFs in the RHIC kinematic region in contrast to the backward (forward)
$\eta_{e}$ region dominated by the behavior of the well-known valence $u$ ($d$) polarized PDFs \cite{deFlorian:2010aa}.
At midrapidity, $W^{+(-)}$ production probes a combination of the polarization of the $u$ and $\bar{d}$ ($d$ and $\bar{u}$) quarks, 
and $A^{W^{+(-)}}_{L}$ is expected to be negative (positive) \cite{deFlorian:2008mr, deFlorian:2009vb}. 
The measured $A^{W^{+}}_{L}$ is indeed negative stressing the direct connection to the $u$ quark polarization. The central value 
of $A^{W^{-}}_{L}$ is positive as expected with a larger statistical uncertainty. Our $A_{L}$ results are consistent with 
predictions using polarized quark and antiquark PDFs constrained by 
inclusive and semi-inclusive pDIS measurements, as expected from the universality of polarized PDFs. 
An independent measurement of $W$ boson production 
from RHIC is being reported by the PHENIX collaboration \cite{Adare:2010xa}.

In summary, 
we report the first measurement of the parity violating single-spin asymmetries for midrapidity, $|\eta_{e}|< 1$, decay positrons and electrons 
from $W^{+}$ and $W^{-}$ boson production in longitudinally polarized $\vec{p}+p$ collisions at $\sqrt{s}=500\,$GeV by the 
STAR experiment at RHIC.
This measurement establishes a new and direct way to explore the spin structure of the 
proton using parity-violating weak interactions in polarized $\vec{p}+p$ collisions. 
The measured asymmetries probe the polarized PDFs at much 
larger scales than in previous and ongoing pDIS experiments and agree well with NLO and resummed calculations using the 
polarized PDFs of DSSV08. 
Future high-statistics measurements at midrapidity together with measurements at forward and 
backward pseudorapidities will focus on constraining the polarization of $\bar{d}$ and $\bar{u}$  quarks.


\begin{acknowledgments}

We thank the RHIC Operations Group and RCF at BNL, the NERSC Center at LBNL and the Open Science Grid consortium 
for providing resources and support. We are grateful to D. de Florian, P. Nadolsky and W. Vogelsang for useful discussions.
This work was supported in part by the Offices of NP and HEP within the 
U.S. DOE Office of Science, the U.S. NSF, the Sloan Foundation, the DFG cluster of excellence `Origin and 
Structure of the Universe' of Germany, CNRS/IN2P3, FAPESP CNPq of Brazil, Ministry of Ed. and Sci. of the 
Russian Federation, NNSFC, CAS, MoST, and MoE of China, GA and MSMT of the Czech Republic, FOM and NWO of 
the Netherlands, DAE, DST, and CSIR of India, Polish Ministry of Sci. and Higher Ed., Korea Research Foundation, 
Ministry of Sci., Ed. and Sports of the Rep. Of Croatia, and RosAtom of Russia.
\end{acknowledgments}


\bibliography{W-PRL-Final-PRL}

\end{document}